\documentclass[usenatbib]{mn2e}
\usepackage{graphics}
\usepackage{amsmath,amssymb}
\usepackage{epsf}
\usepackage{dcolumn}
\usepackage{color}
\usepackage{epsfig}
\usepackage{psfrag}
\input{epsf}

\def \beq  {\begin{equation}}
\def \eeq  {\end{equation}}
\def \ber  {\begin{eqnarray}}
\def \eer  {\end{eqnarray}}

\newcommand\keV{\mbox{keV}}

\begin{document}
\title[Thawing Dark Energy using Cluster Counts]
{Constraining Thawing Dark Energy using Galaxy Cluster Number Counts}
\author[Devi, Choudhury \& Sen]
{N. Chandrachani Devi$^{1,2}$\thanks{E-mail:chandrachani@gmail.com},~
T. Roy Choudhury$^{3,4}$\thanks{E-mail: tirth@ncra.tifr.res.in},~
and
Anjan A Sen$^1$\thanks{E-mail:aasen@jmi.ac.in}
 \\
$^1$Center For Theoretical Physics, Jamia Millia Islamia, New Delhi 110025, India\\
$^2$Departamento de Astronomia, Observatorio Nacional, 20921-400, Rio de Janeiro - RJ, Brasil\\
$^3$Harish-Chandra Research Institute, Chhatnag Road, Jhusi, Allahabad 211019, India\\
$^4$National Centre for Radio Astrophysics, TIFR, Post Bag 3, Ganeshkhind, Pune 411007, India
}

\date{\today}

\maketitle

\begin{abstract}
 We study the formation of galaxy clusters in the presence of  thawing class of scalar field dark energy.  We consider cases where the scalar field has canonical as well non canonical kinetic term in its action. We also consider various form for the potential of the scalar field e.g, linear, quadratic, inverse quadratic, exponential as well as Pseudo-Nambu-Goldstone Boson (PNGB) type. Moreover we investigate situation where dark energy is homogeneous as well as the situation where dark energy takes part in  virialization process. We use the Sheth-Tormen formalism while calculating the number density of galaxy clusters. Our results show that cluster number density for different dark energy models have significant deviation from the corresponding value for the $\Lambda$CDM case. The deviation is more for higher redshifts. Moreover the tachyon type scalar field with linear potential has the highest deviation from  the $\Lambda$CDM case. For the total cluster number counts, different dark energy models can have substantial deviation from $\Lambda$CDM and this deviation is most significant around $z \sim 0.5$ for all the models we considered. We also constrain thawing class of models using the presently available data for number counts of massive X-ray clusters. The results show that current cluster data is not suitable enough for constraining potentials for the thawing scalar fields as well as for other cosmological parameters like $n_{s}$. But one can get significant constraint for the parameter $\sigma_{8}$ and a lower bound on $\Omega_{m0}$.

\end{abstract}

\section{Introduction}

Over the last decade, the observational data from  Supernovae Type Ia (SNIa) \citep{Kowalski2008,Riess2009}, Cosmic Microwave Background Radiation (CMBR)\citep{Komatsu2011}, Baryon Acoustic Oscillations (BAO)\citep{Percival2009} and the large scale structure surveys
\citep{Cole2005} have confirmed that our Universe at present is going through an accelerated expanding phase. Till date, there has been a large number of proposals to explain such an accelerated expansion. This includes the inclusion of an unknown homogeneous matter component having a large negative pressure (cosmological constant being the simplest example of such fluid), modification of gravity at large scale as well as considering the back-reaction of small scale inhomogeneities in the matter distribution.  

Although inclusion of cosmological constant in the energy budget of the universe is a minimal way to explain the late time cosmological acceleration which is also allowed by all cosmological observations, but at the same time it is plagued with the fine tuning and the cosmic coincidence problems \citep{Weinberg1986, Sahni2000, Carroll2001}. 
Scalar field models with generic features can alleviate these problems and provide an alternative to cosmological constant. These dynamical scalar field models of dark energy are broadly classified into two categories: fast roll and slow roll models dubbed freezing and thawing models. For  details,  see the reference \citep{Caldwell2005}. Among these, thawing scalar field models are particularly interesting as they can naturally mimic equation of state very close to $w \sim -1$ which is preferred by all the observational data.

On the other hand, information about the abundance of collapsed structures as a function of mass and redshift is an important tool to study the matter distribution in the universe \citep{Evrard2002}. A large number of cluster surveys are ongoing or 
being planned to be set-up in near future, e.g., 
PLANCK, eROSITA, WFXT which would detect a large number of clusters \citep{Vikhlinin2009b}. Indeed, the mass functions of galaxy clusters have been measured through X-ray surveys \citep{Borgani2001,Reiprich2002,Vikhlinin2009a}, via weak and strong lensing studies \citep{Bartelmann1998,Dahle2006,Corless2009}, using optical surveys, like the SDSS \citep{Bahcall2003a,Wen2010} as well as through Sunayev-Zeldovich effect \citep{Tauber2005}. In the last decade several authors have been involved in such studies and have found that the dark energy not only affects the expansion rate of the background and the distance-redshift relation but also the growth of structure in the
Universe \citep{Weinberg2003, Wang1998, Manera2006, Liberto2006, Nunes2006, Francis2009, Mortonson2009, Pace2010, Khedekar2010a, Khedekar2010b, basik2009, basik2010}. The investigation by \citet{basik2010} is  particularly interesting in this regard. They have analyzed the predicted cluster-size halo redshift distribution on the bass of two cluster surveys: one is the future X-ray survey based on the e-Rosita satellite and the other one is the Sunayev-Zeldovich survey based on the South Pole Telescope. They found that the predictions of most of the popular dark energy models can be clearly distinguished from the concordance $\Lambda$CDM model based on the observations that one can have from these two surveys.
Hence dark energy is expected to have an impact on observables such as cluster number counts and lensing statistics \citep{Evrard2002}. Therefore, the studies of galaxy clusters would provide a useful tool to constrain the model parameters and would help to infer the properties of dark energy by discriminating among the different dark energy models.

Although the proper way to understand the effect of dark energy in the non-linear regime of structure formation is through the N-body numerical simulations \citep{Baldi2010, Courtin2011, Maccio2004}, one can also use a simpler semi-analytical methods which are in reasonable agreement with N-body simulations.  One such example is the spherical collapse model \citep{Gunn1972} using the Press-Schechter formalism\citep{Press1974}. To overcome the problem of over-prediction of number of low(high) mass halos at the current epoch for the above model, an elliptical collapse model was later proposed by \citet{Sheth1999}.

This paper extends the previous work by \citet{Devi2011} by further studying the most general analytic form of halo mass function introduced by \citet{Sheth1999} for thawing type scalar field dark energy models with various potential and examine how  does the DE affect the abundance of CDM halos by measuring the galaxy cluster number counts. We consider both the canonical and non-canonical form of thawing dark energy models. For the completeness of the study we check the results for homogeneous as well as inhomogeneous dark energy cases. We discuss how much they deviate from the conventional $\Lambda$CDM model through the linearly extrapolated density contrast $\delta_{c}(z)$ at the redshift of collapse, the number density of halo mass at some particular redshifts of collapse and also through the total cluster number counts.

In a recent paper \citet{Campanelli2011} have studied the constraints on different dark energy models through cluster number counts. They assumed the CPL parametrization \citep{Linder2003, Chevallier2001} to model different dark energy models. But it was recently shown by \citet{Gupta2012} that this parametrization does not correctly represent all the thawing class of scalar field models. Moreover this parametrization is also not suitable for thawing scalar fields with non canonical kinetic term. Hence in our analysis, we do not assume any parametrization that represents thawing class of models. Instead, we consider the full system of coupled scalar field plus Einstein's equations for the thawing class of scalar fields.

We also put constraints on our model using the presently available cluster number data from massive X-ray clusters as obtained by \cite{Campanelli2011}. In doing this we concentrate on canonical scalar field models that deviate significantly from $\Lambda$CDM behaviour. The results shows that present data is not suitable to constrain the thawing scalar field potentials as well as other cosmological parameters like $n_{s}$, although one can get significant constrain on the parameter $\sigma_{8}$, the rms mass fluctuation today at $8 h^{-1} Mpc$ scale and a lower bound on $\Omega_{m0}$.

The structure of the paper is as follows: in section 2, we introduce the thawing dark energy models for both canonical and non-canonical kinetic term. We discuss the background evolution for such fields considering the difference type of potentials e.i. $V=\phi$, ~$V = \phi^2$, $V=e^\phi$ and $V = \phi^{-2}$  and also PNGB type. In section 3, we sketch the derivation of the equations to calculate the linearly extrapolated density contrast, $\delta_c(z)$, at the  collapsed redshift. Then, we describe the halos mass function introduced by \citet{Sheth1999} and calculate the  number density and the total cluster number counts for our DE models. 
In section 4, we discuss the observational constrain on the thawing models and finally we draw conclusions in section 5.

\section{Background Evolution}

In what follows, we consider a flat, homogeneous and isotropic background universe driven by non-relativistic matter and dark energy of thawing type, i.e. $\Omega_\phi + \Omega_m = 1$. These thawing type dark energy models are characterized by the fact that in the early universe the scalar field is frozen by very large Hubble damping and the scalar field starts evolving slowly down its potential at the later time. So, the equation of state,  ~$w(a)=p_{\phi}/\rho_{\phi}$ initially starts with $w=-1$ and slowly departs from it in the later time. We consider both ordinary scalar field with canonical kinetic term as well as tachyon type scalar field having Born-Infeld type kinetic term which are minimally coupled to the gravity sector \citep{Sen2002a,Sen2002b,Garousi2000,Kluson2000}.
The equation of motions for the canonical scalar field and the tachyon field are given by

\begin{eqnarray}
\ddot{\phi}+3 H{\phi}+ \frac{dV}{d\phi} = 0 
\label{fieldcanonical}~~~{\rm and}\\
~~~\ddot{\phi}+3H\dot{\phi}(1-\dot{\phi}^2)+\frac{dV/d\phi}{V}(1-\dot{\phi}^2)=0
\label{fieldnoncanonical}
\end{eqnarray}
respectively, where dot represents the differentiation w. r. t the cosmic time $t$ and the Hubble parameter, $H$ is defined as
\begin{equation}
H^2 = \left(\frac{\dot{a}}{a}\right)^2 = \frac{8 \pi G}{3}({\rho_m}+{\rho_{\phi}}).
\label{H}
\end{equation}
Here $a(t)$ is the scale factor, $\rho_m=\rho_{m0}a^{-3}$ is the background matter density and $\rho_{\phi}=\rho_{\phi0}{f(a)}$ represents the dark energy density with 
\begin{equation}
 {f(a)}= \exp\left[3\int_{a}^{1}\left(\frac{{1+w(u)}}{u}\right) {\rm d}u\right]. 
\label{f}
\end{equation}
Defining new variables $\lambda \equiv -\frac{1}{V}\frac{dV}{d\phi}$ and $\Gamma \equiv V \frac{d^2 V}{d\phi^2}/\left(\frac{dV}{d\phi} \right)^2$ for the canonical scalar field model, one can form an autonomous system
of equations involving two observable parameters $\Omega_{\phi}$ and $\gamma = (1+w)$ together with the parameter $\lambda$.  For the canonical scalar field, it is given as
\begin{eqnarray}
\gamma^\prime &=& -3\gamma(2-\gamma) + \lambda(2-\gamma)\sqrt{3 \gamma
\Omega_\phi},
\label{Omegaprime}\\
\Omega_\phi^\prime &=& 3(1-\gamma)\Omega_\phi(1-\Omega_\phi),
\label{lambda}\\
\lambda^\prime &=& - \sqrt{3}\lambda^2(\Gamma-1)\sqrt{\gamma \Omega_\phi}.
\label{autonomous1}
\end{eqnarray}
For detail derivations of the above equations, see \cite{Scherrer2008a}.
Here the prime denotes  derivative with respect to ln$a$. We solve this system of equations numerically for the mentioned potentials providing the initial conditions for $\gamma$ and $\lambda$ as $\gamma = 0$, $\lambda_{i} = 1$. As discussed in \citet{Scherrer2008a}, one can easily see that smaller the value of $\lambda_{i}$, more the scalar field evolution similar to that of cosmological constant $\Lambda$. To check for the maximum deviation from the cosmological constant $\Lambda$, we set $\lambda_{i} = 1$.
 For $\Omega_{\phi}$ we choose its initial value in such away so that to get required value at present.
The different types of potential we considered are $V=V_{0}\phi$, ~$V =V_{0} \phi^2$, $V=V_{0}e^\phi$ and $V = V_{0}\phi^{-2}$ and they correspond to $\Gamma=0,~\frac{1}{2},~1$ and $\frac{3}{2}$ respectively. We also considered the Pseudo-Nambu Goldstone Boson model (PNGB) \citep{Frieman1995} which has been characterized by the potential:
\begin{equation}
V(\phi) = m^4 \left[\cos\left(\frac{\phi}{f}\right)+1\right] 
\end{equation}
with $f=1$. To solve the above of system of equations, one needs a third initial condition for $\Omega_{\phi}$. The parameters for the potentials, $V_{0}$ or $m$ can be related to this initial value for $\Omega_{\phi}$. Moreover this initial value for $\Omega_{\phi}$ can be related to its value at present, $\Omega_{\phi 0}$. Hence once we choose the form of the potentials and the initial values for $\gamma$ and $\lambda$ as mentioned above, the only free parameter for the background evolution is $\Omega_{\phi 0}$ or $\Omega_{m0}$. The rest of the potential parameters can be known from its value.

\begin{figure}
\includegraphics[width=80mm]{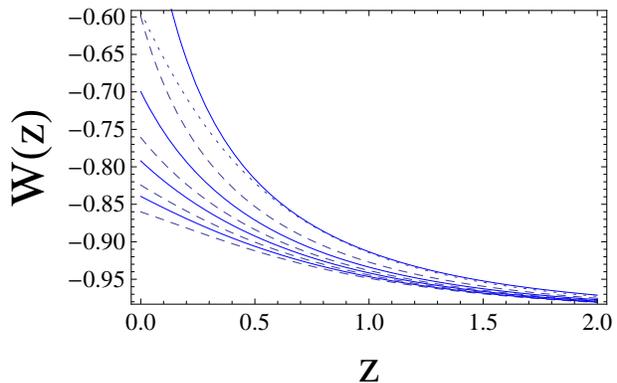}
\caption{Evolution of the equation of state $w$ for different scalar field
and tachyon models.
Solid curves represent different Tachyon models with $V(\phi) = \phi, \phi^2, e^{\phi}, \phi^{-2}$ respectively from top to bottom,
Dashed curves from top to bottom represent different scalar field models with same
potentials as in tachyon. Dotted curve represents PNGB model. $\Omega_{m0} = 0.25$.}
\label{fig:wz}
\end{figure}

Similarly, for the thawing tachyon field models, the autonomous system of equations is given by
\begin{eqnarray}
\gamma'&=&-6\gamma(1-\gamma)+2\sqrt{3\gamma\Omega_{\phi}}\lambda(1-\gamma)^\frac{5}{4}
\label{tachgamma}\\
\Omega_{\phi}'&=&3\Omega_{\phi}(1-\gamma)(1-\Omega_{\phi})
\label{tachomega}\\
\lambda'&=&-\sqrt{3\gamma\Omega_{\phi}}\lambda^2(1-\gamma)^\frac{1}{4}(\Gamma-\frac{3}{2})
\label{autonomoustach}
\end{eqnarray}
where $\lambda=-\frac{1}{V^{3/2}}\frac{dV}{d\phi}$ and $\Gamma=V\frac{d^{2}V}{d\phi^2}/\frac{dV}{d\phi}$. Here also we set the initial conditions similar to the canonical scalar field case and consider the power law potentials as mentioned above. For the detail calculations see \cite{Scherrer2008a,Scherrer2008b,Amna2009}. One can see the behaviour of the equation of state parameter $w(z)$ from the Figure \ref{fig:wz} for the various models we considered. The different models have a maximum deviation when one approaches the
present day however they behave almost identically in the past. 

\section{Halo Mass Function}

The studies of the mass function of collapsed objects like 
clusters of galaxies can be used to constrain the cosmological models and can help to infer the properties of dark energy. In what follows we calculate the halo mass function for the DE models considered in this paper.

\subsection{Spherical collapse}

As our interests lie in finding the halo-mass of CDM matter in presence of thawing dark energy, we need to study the perturbation  of matter inhomogeneity in order to calculate the linear density contrast at the time of collapse. 
We refer the readers to \citet{mota2008} and \citet{basik2009} for some earlier works regarding spherical collapse formalism with dark energy models.
We consider a spherical region of radius $r(t)$ evolving in a
cosmologically expanding background.
The dynamics of this spherical region is essentially 
governed by the Raychaudhuri equation,
\begin {equation} 
\frac{\ddot{r}}{r}=-4 \pi G\left[ \left(w(r)+\frac{1}{3}\right) {\rho_{\phi cl}}+\frac{1}{3}{\rho_{m cl}}\right] 
p\label{eq:raychaudh}
\end{equation}
where $\rho_{\phi cl}$ and $\rho_{m cl}$ are the density of scalar field and the matter density inside the cluster respectively.
It is easier to solve the equations after normalizing at the turn around point, so we define new variables: 
\begin{equation}
  x=\frac{a}{a_t}\;{\rm and}\;y=\frac{r}{r_t}.
\end{equation}
where the subscript $t$ denotes the turn around time. Now, the equations of background evolution and that of perturbation reduce to
\begin{equation}
  \dot{x}^2={H_t}^2\Omega_{m,t}[ \Omega_m(x)x]^{-1}
\label{eq:background} 
\end{equation}
 and
\begin{equation}
{\ddot y}=-\frac{H_{\rm t}^{2}\Omega_{\rm m,t}}{2}
\left[ \frac{\zeta}{y^{2}}+\nu y I(x,y)\right] \;\;\;
\label{eq:perturbation}
\end{equation} 
where the function $I(x,y)$ is given by \citep{Basilakos2007}
\begin{equation}
I(x,y)=\left\{ \begin{array}{cc}
       \displaystyle\left[1+3w(r(y))\right]\frac{f(r(y))}{f(a_t)} &
       \mbox{Clustered DE}\\
       \left[1+3w(x)\right]f(x) & \mbox{Homogeneous DE}
       \end{array}
        \right.
\label{eq:I}
\end{equation}
with
\begin{equation}
\nu=\frac{\rho_{\phi, t}}{\rho_{m, t}}=\frac{1-\Omega_{m,
    t}}{\Omega_{m, t}}.
\end{equation}
Here $\zeta$ represents the matter density contrast at turnaround which is defined as
 \begin{equation}
\zeta \equiv\frac{\rho_{mcl,t}}{\rho_{m,t}}=\left(\frac{R_t}{a_t}\right) ^{-3}.
 \end{equation}
 The function $r(y)$ is given by $r(y)=r_{\rm t}y=\zeta^{-1/3} a_{\rm t}y$ and $\Omega_{m} (x)$ by
\begin{equation}
\Omega_m(x)=\frac{1}{1+\nu x^{3} f(x)} .
\end{equation}
We assume here that the equation of dark energy $w$ has the same form inside and outside the cluster. This is purely an assumption to simplify the analysis and is not necessarily true. But one should note that the $w$ is function of the background scale factor $a$ outside the cluster, but inside the cluster, it is a function of the radius $r(t)$ of the spherical region. Hence the actual behavior of the equation of state $w$ with time will be different inside and outside the over density, which is similar to the work by \citet{basik2009}.

\begin{figure*}
\vspace{1.5cm}
\psfrag{z}[c][c][1][0]{{\bf {\Large $z_c$}}}
\psfrag{d}[c][c][1][0]{{\bf {\Large $\delta_c$}}}
\includegraphics[width=14cm]{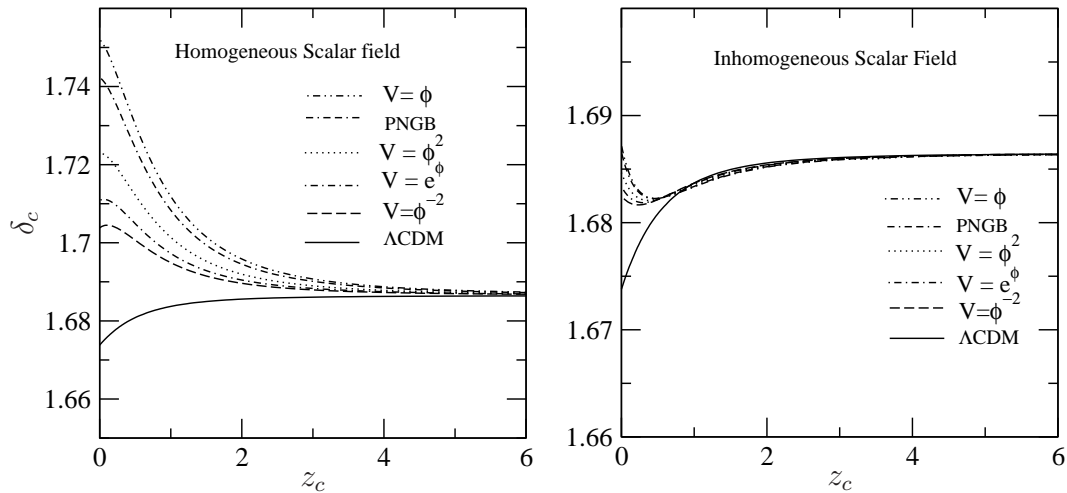}
\caption{The redshift evolution of the linear density contrast,  $\delta_{c}$ at the redshift of collapse for the canonical scalar field model with various potentials for $\Omega_{m0} = 0.25$. {\it Left panel}: Homogeneous dark energy models. {\it Right panel}: Inhomogeneous dark energy models. In all panels, the different potentials $V(\phi)=\phi, \phi^2, e^{\phi}$ and $\phi^{-2}$ are represented by different line types along with PNGB model. The $\Lambda$CDM case (black solid curve) is plotted for reference.  At high redshift, all the models asymptotically approach to the Einstein-de Sitter limit.}
\label{fig1:lindenscalar}
\end{figure*}

In addition, the linear density contrast $\delta$ obeys the equation:
\begin{equation}
{\ddot{\delta}}  +  2\frac{\dot{a}}{a}{\dot{\delta}}
  = 4\pi G \rho_{m}\delta 
  = \frac{3}{2}H_{0}^2\Omega_{m0}a^{-3}\delta .
\label{eq:lineardensity}
\end{equation}
We calculate the linear over density $\delta_{c}$ at  the 
epoch when the spherical region described by equation (\ref{eq:raychaudh}) 
collapses to a point, $t(z_{c}) = 2 t(z_{t})$ by solving equation (\ref{eq:lineardensity}). Here we assume that the virialization epoch is approximately twice of the turn around epoch.
We should mention that this condition may get violated in some cases, like a dark energy model with phantom equation of state. But as we are considering here scalar field models,
we never go to the phantom region of the equation of state. Hence this relation still holds for our case. For the initial condition 
$\delta_{i}=\left(\rho_{mcl}/\rho_{m}-1\right)_{a\rightarrow 0}$, we write
\beq
\left[\frac{r}{a}\right]_{a\rightarrow 0} =\left[\zeta^{-1/3}\frac{y}{x}\right]_{x\rightarrow 0} = (1-\beta a_{t}x)
\label{eq:ra}
\eeq 
so that,
\beq
\left[\frac{dy}{dx}\right]_{x\rightarrow 0} = \zeta^{1/3}(1-\beta a_{t}x).
\label{eq:dydx}
\eeq 
Here $\beta$ is the constant term which can be found by solving equations (\ref{eq:background}) and (\ref{eq:perturbation}) after substituting equations (\ref{eq:ra}) and (\ref{eq:dydx}). We neglect the higher order terms in $x$. For our case the initial conditions on $\delta_{i}$  are found out as
\begin{equation}
\delta_{i}
=\left\{ \begin{array}{cc}
       \displaystyle\left[\zeta^{1/3}+\zeta^{-2/3}\nu \frac{f(r(1))}{f(a_{t})}\right]\frac{a_{i}}{a_{t}} &
       \mbox{Clustered DE}\\
       \displaystyle\left[\zeta^{1/3}+\zeta^{-2/3}\nu f(1) \right]\frac{a_{i}}{a_{t}} & \mbox{Homogeneous DE}
       \end{array}
        \right.
\end{equation}
As we are dealing with second-order equations, two initial values are required, one for the initial over-density $\delta_{i}$
 and other is the initial rate of evolution, $\delta^{'}_{i}$. The quantity 
$\delta^{'}_{i}$ is generally set to $ \delta^{'}_{i}=10^{-5}$. However we set  $\delta^{'}_{i}=0$ since the result does not change considerably. The initial 
epoch $a_{i}$ is set to $ a_{i} = 10^{-3}$.

\subsection{Number counts}

With the indication from observations that individual galaxies and cluster of galaxies are embedded in extended halos of dark matter, the abundance of CDM halos have been studied widely. Theoretically, Press and Schetcher were the first to describe the abundance of these CDM halos as a function of their mass with the assumption that the fraction of the volume of the universe that has collapsed into objects of mass $M$ at a redshift $z$ follows a Gaussian distribution.
The comoving number density of clusters which have collapsed 
(i.e., virialized) at certain redshift $z$ and have masses in the range $M \sim M+dM$ can be expressed as:
\beq
\frac{dn(M,z)}{dM} =-\frac{\rho_{m0}}{M}\frac{d \ln \sigma(M,z)}{dM}f(\sigma(z))
\label{dndm}
\eeq
where $f(\sigma(z))$ is defined as mass function. The standard Press-shechter mass function is of the form :
\beq
f(\sigma ; PS) = \sqrt{\frac{2}{\pi}}\frac{\delta_c(z)}{\sigma(M,z)}\exp\left[- \frac{\delta_c^2(z)}{2\sigma^2(M,z)}\right].
\eeq
Although it provides a good general representation of the observed distribution of clusters, due to the discrepancy of over-prediction (under-prediction) of the number of low (high) mass halos at the current epoch, \citet{Sheth1999} introduced an ellipsoidal model of the collapse of perturbations. This Sheth-Tormen (ST) mass function gives better fits to simulated mass function by reducing this discrepancy substantially. It has the form:
\beq
f(\sigma ; ST) = A\sqrt{\frac{2a}{\pi}}\left[1+\left(\frac{\sigma^2}{a\delta_c^2(z)}\right)^p\right]\frac{\delta_c(z)}{\sigma}\exp\left[- \frac{\delta_c^2(z)a}{2\sigma^2}\right];
\eeq
it contains three parameters $A, a$ and $p$ which we set into $ A=0.322, a=0.707$ and $p=0.3$ for all the models we considered. The Press-Shechter case is recovered for $a=1$ and $p=0$.

It is worth mentioning here that the accuracy of the ST mass function has been found to be limited. In fact,
recent high-quality N-body simulations
have been used to find fitting functions which work far better than ST mass function for varying dark energy models
\citep{basik2010,Bhattacharya2011}. However, we would keep working with the
ST mass function in this paper, keeping in mind that this may introduce
$\sim 10\%$ errors in our calculations \citep{Reed2007}. Since we are only
studying the viability of using future surveys in distinguishing between
dark energy models, this much error can be tolerated at this stage. 
In case one wants
to study, e.g.,  precision cosmology with future cluster data, a much more
sophisticated fit would be mandatory.

\begin{figure*}
\vspace{1.5cm}
\psfrag{z}[c][c][1][0]{{\bf {\Large $z_c$}}}
\psfrag{d}[c][c][1][0]{{\bf {\Large $\delta_c$}}}
\includegraphics[width=14cm]{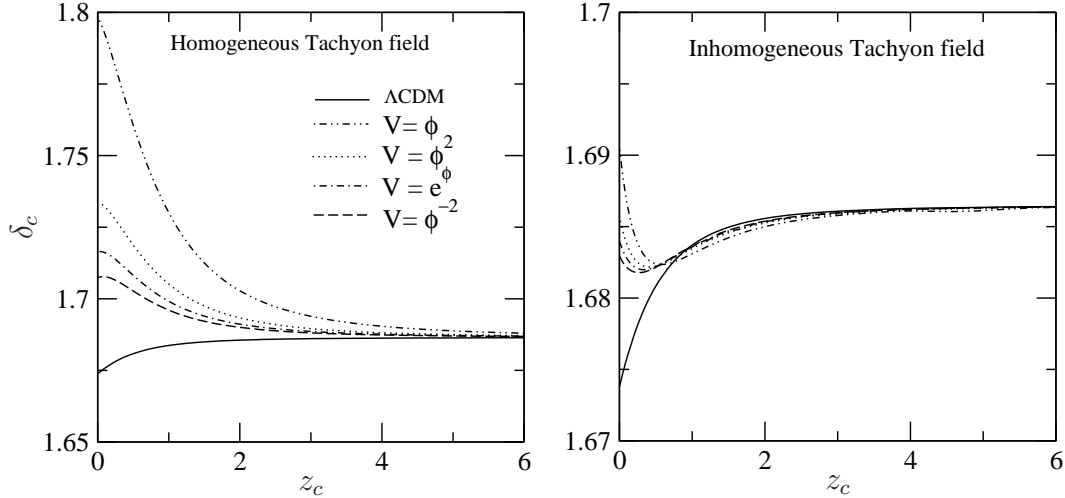}
\caption{Same as figure \ref{fig1:lindenscalar} but for the tachyon dark energy models. The various potentials are indicated by different line types. All the models asymptotically approach to the EDS limit at hight redshift.}
\label{fig2:lindentach}
\end{figure*}

\begin{table*}
\centering
\begin{tabular}{|c|c|c|c|c|c|}

\hline
Model & Potential & Case & $a_{1}$ & $b$ & $c$  \\

\hline 
 & $ \phi $  & inhom & $1.100 \pm 0.001$ & $0.153 \pm 0.001$ & $0.0607 \pm 0.0003 $\\   
       &          & hom & $ 0.8972 \pm 0.0005$ & $- 0.04884 \pm 0.0005$ & $-0.0913 \pm 0.0003$\\
\cline{2-6}      
 & $ \phi^{2}$ & inhom & $ 1.100 \pm 0.001$ & $ 0.154 \pm 0.001$ & $0.0607 \pm 0.0004$ \\
       &          & hom & $0.662 \pm 0.004$ & $- 0.284 \pm 0.004$ & $- 0.143 \pm 0.002$ \\
       \cline{2-6}
Scalar&  $\phi^{-2}$ & inhom & $ 0.8619\pm 0.0007 $ & $-0.0841 \pm 0.0007 $& $-0.0154 \pm 0.0002$\\
       &          & hom & $ 1.135 \pm 0.005$ & $0.189 \pm 0.005 $ & $0.026 \pm 0.001 $\\
       \cline{2-6}
& $ e^{\phi} $ & inhom & $0.729 \pm 0.006$ & $-0.217 \pm 0.006$ & $-0.0591 \pm 0.0002$ \\
      &           & hom & $1.399 \pm 0.008$ & $ 0.453 \pm 0.008$ & $ 0.106 \pm 0.002$\\
     
\hline 
\end{tabular} 
\caption{Values of the fitting parameters $a_{1}, b$ and $c$ by fitting equation (\ref{fitdeltac}) of density contrast at the collapse point as a function of $\Omega_{m}$ and $z_c$ for the canonical scalar field model with some specific potentials. The errors at $95\%$ confidence level are also shown.} 
\label{tab:fitting}
\end{table*}

The dispersion of the density field on a given comoving scale $R$, containing mass $M=4\pi\rho_{m0}R^3/3$, is given by
\beq
\sigma^2(R) = \frac{D(a)^2}{2\pi^2}\int_0^{\infty} k^3 P(k)W^2(kR)\frac{dk}{k}
\eeq
where the quantity $P(k)$ is the power spectrum of density fluctuations extrapolated to $z = 0$ according to linear theory and $W(kR)$ is  the top-hat window function;
\beq
W(kR) = 3 \left(\frac{sin(kR)}{(kR)^3}-\frac{cos(kR)}{(kR)^2}\right).
\eeq
 $D(a)$ represents the growth function of linear perturbation theory and can be found from equation(\ref{eq:lineardensity}) (see fig.4 in \citet{Schaefer2008}).
We normalize the growth function such that $D(a)=1$ at the present epoch. 

Assuming that the baryon density parameter $\Omega_{B0}\ll\Omega_{CDM,0}$, the CDM power spectrum can be approximated by $P(k) = P_0 k^{n_s} T^2(k)$, where $P_0$ is
a normalization constant and we use the transfer function, $T(k)$ introduced by Eisenstein and Hu \citep{Eisenstein1998a}. 

The normalization of the power spectrum $P_0$ is often expressed in terms of $\sigma_8$, the rms fluctuation today at a scale of 8 $h^{-1}$Mpc.
However, since the differences between various dark energy models is
usually most prominent at lower redshifts, normalizing the power spectrum using
$\sigma_8$ could lead to erroneous results. We rather normalize using
the observed CMBR power spectrum amplitude \citep{basik2010}:
\beq
P_0 \simeq 2.2 \times 10^{-9} \left(\frac{2}{5 \Omega_{m0}}\right)^2
H_0^{-4} k_0^{1-n_s}
\eeq
where $k_0 = 0.02$ Mpc$^{-1}$ is a characteristic length probed by CMBR
observations \citep{Komatsu2009}.
   
The number of clusters in a redshift interval $dz$, above a given minimum
(threshold) mass $M= M_{min}$ is obtained from $dn(M,z)/dM$:
\begin{equation}
N(M>M_{min},z) = f_{sky}\frac{dV(z)}{dz}\int_{M_{min}}^{\infty} dM\frac{dn}{dM}(M,z)
\label{eq:dNdz}
\end{equation}
where $f_{sky}$ is the fraction of the sky being observed and the comoving volume element is given by 
\begin{equation}
\frac{dV}{dz}= 4\pi r^2(z)\frac{dr}{dz}.
\end{equation}
$r(z)$ is the comoving radial distance out to redshift z:
\beq
r(z)= c\int_0^z \frac{dz^{'}}{H(z^{'})},
\eeq
where $H(z)$ is the Hubble parameter. For numerical computation, the upper limit of integration in equation (\ref{eq:dNdz}) is replaced by some finite mass value $M_{max}$. The comoving volume element is required since the redshift evolution of a physical volume in space is model dependent, i.e, depending on the form of potentials. 

\subsection{Comparison of Thawing Models with $\Lambda$CDM}

In this section, we discuss the results for the linear over-density contrast, the number density of CDM halos and the total clusters number counts for the models we introduced, keeping the $\Lambda$CDM model as a reference since the $\Lambda$CDM model is 
currently the simplest model, fitting all available observational data despite of its conceptual problems.

In Fig.\ref{fig1:lindenscalar}, we show the linear density contrast, $\delta_c$, as a function of  collapsed redshift for  both homogeneous (left panel) and inhomogeneous (right panel) scalar field models with various potentials. Similar behavior is being shown for the case of tachyon field in Fig.\ref{fig2:lindentach}. From Figs.\ref{fig1:lindenscalar} and \ref{fig2:lindentach}, it can be seen that the linear density contrast $\delta_{c}$ at $z=0$  has a significant deviation from the $\Lambda$CDM case for homogeneous dark energy models.  These deviations are comparatively smaller in case of inhomogeneous dark energy. This is true for both ordinary scalar field as well as tachyon type scalar field. This type of behavior is expected. For homogeneous case, the equation of state of dark energy is greater than $w=-1$ (we are not considering Phantom models). Hence the repulsive effect of dark energy in the background evolution is lesser than the $\Lambda$CDM case. This results larger linear density contrast for homogeneous dark energy. However, for inhomogeneous case, there is an extra repulsive effect inside the cluster due to inhomogeneous dark energy. This reduces the linear density contrast for inhomogeneous DE and brings it closer to the $\Lambda$CDM value. At high redshift, all the models asymptotically approach to the Einstein-de Sitter (EDS) limit. Among the potentials we considered, the linear potential shows maximum  deviation from the $\Lambda$CDM for  the ordinary and tachyon fields in both the homogeneous and inhomogeneous cases. This is consistent with the results earlier obtained by \citet{Devi2011}. 

\begin{figure*}
\centering
\begin{center}
$\begin{array}{@{\hspace{-0.10in}}c@{\hspace{0.0in}}c}
\multicolumn{1}{l}{\mbox{}} & \multicolumn{1}{l}{\mbox{}} \\
[-0.2in] \epsfxsize=3.3in \epsffile{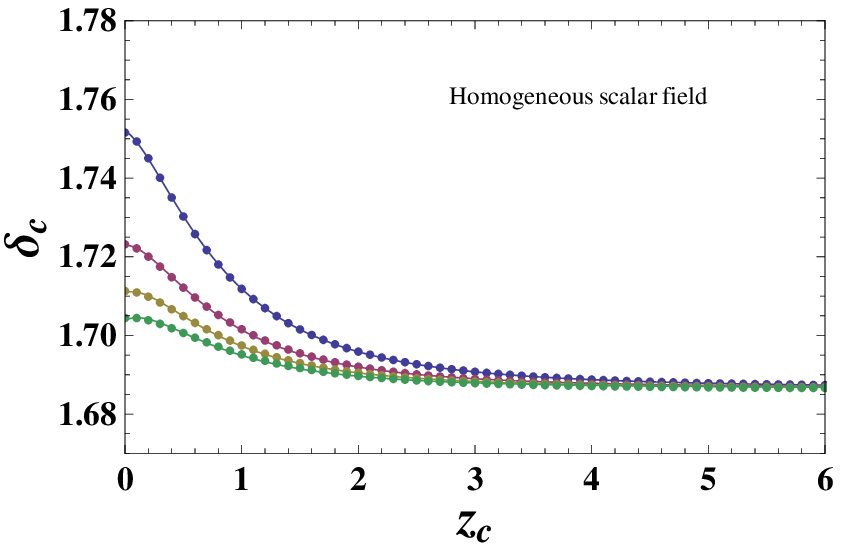} & \epsfxsize=3.3in
\epsffile{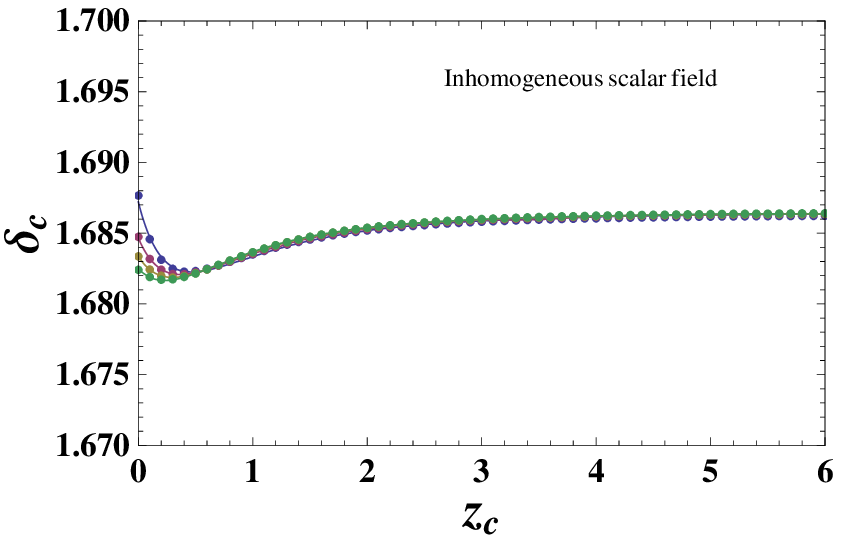} \\
\end{array}$
\end{center}
\vspace{0.0cm} 
\caption{\small 
Plot of the fitting function for the extrapolated linear density contrast at collapse point $\delta_{c}$ vs the collapse redshift ($z_{c}$) for different models. From top to bottom represents scalar field having canonical kinetic term with linear, $\phi^2$, exponential and $\phi^{-2}$ potentials. For each case, the smooth line is for the fitting equation given by eq.(\ref{fitdeltac}) with the parameters mentioned in table \ref{tab:fitting}, whereas the dots are generated by the exact numerical result for $\delta_{c}$ calculated.} 
\label{fig:fitting} 
\end{figure*}
\begin{figure*}
\vspace{1.73cm}
\psfrag{M}[c][c][1][0]{{\bf {\Large ${\rm Log_{10}}(M/h^{-1}M_{\odot})$}}}
\psfrag{alpha}[c][c][1][0]{{\bf {\Large $\alpha$}}}
\includegraphics[width=14cm,height=12cm]{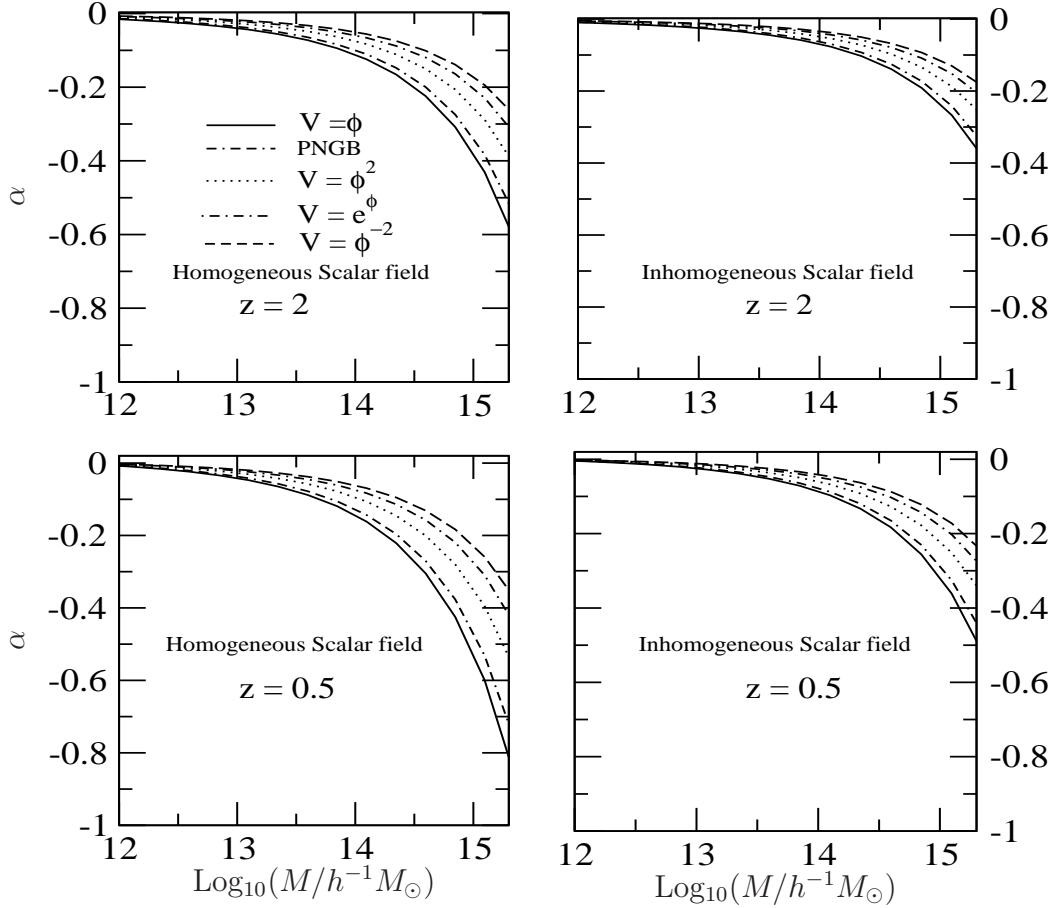}
\caption{A comparison between the comoving number density of CDM halos for the different thawing scalar field models with that of  $\Lambda$CDM model by defining a parameter, $\alpha$ in equation \ref{eq:alpha}. {\it Upper panels}: Homogeneous and inhomogeneous dark energy models at redshift, $z=2$ in the left and right panels respectively. {\it Lower panels}: Homogeneous and inhomogeneous dark energy models at redshift, $z=0.5$ in the left and right panels respectively.
In all panels, the various potentials are indicated by different line types. 
The cosmological 
parameters are taken to be
$\Omega_{m0}=0.25,~
\Omega_{\phi0}=0.75,~
h = 0.702,~
\Omega_{B0} = 0.0456~{\rm and}~
n_s= 0.968$. The normalization of the power spectrum is mentioned in the text.}
\label{fig3:diffscalar}
\end{figure*}  

\begin{figure*}
\vspace*{1.5cm}
\psfrag{M}[c][c][1][0]{{\bf {\Large ${\rm Log_{10}}(M/h^{-1}M_{\odot})$}}}
\psfrag{alpha}[c][c][1][0]{{\bf {\Large $\alpha$}}}
\includegraphics[width=14cm,height=12cm]{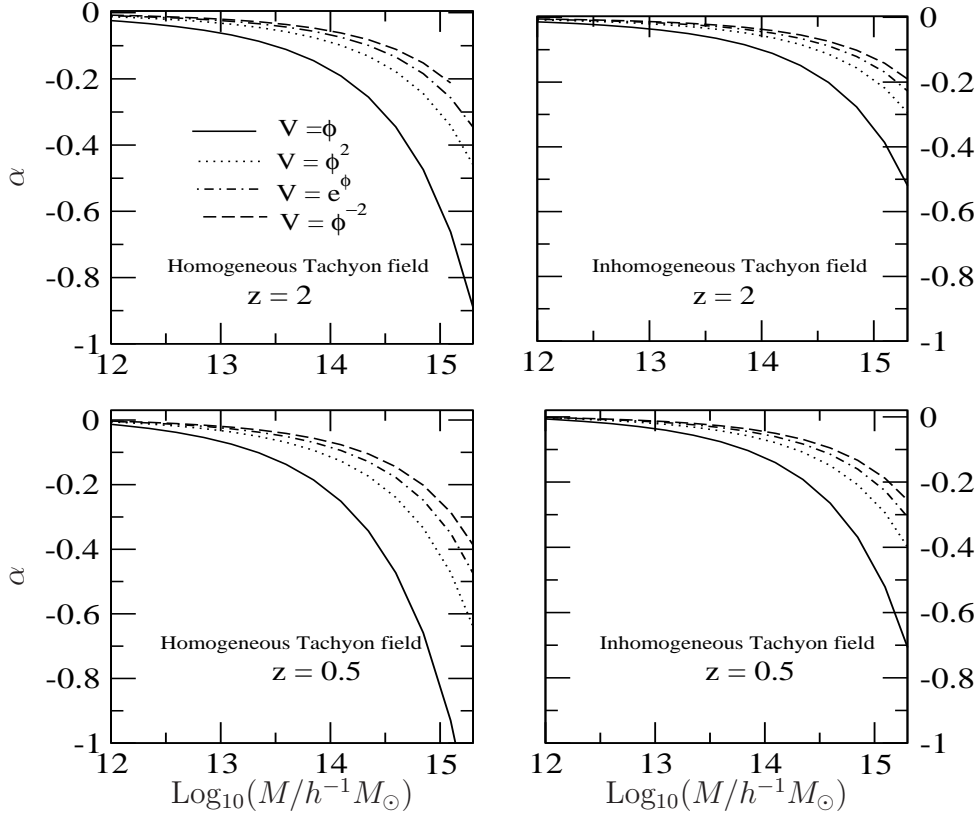} 
\caption{Same as figure \ref{fig3:diffscalar}. for the tachyon field dark energy models. The various potentials are indicated by different line types.}
\label{fig4:difftach}
\end{figure*}

\begin{figure*}
\vspace{1.5cm}
\includegraphics[width=14cm,height=12cm]{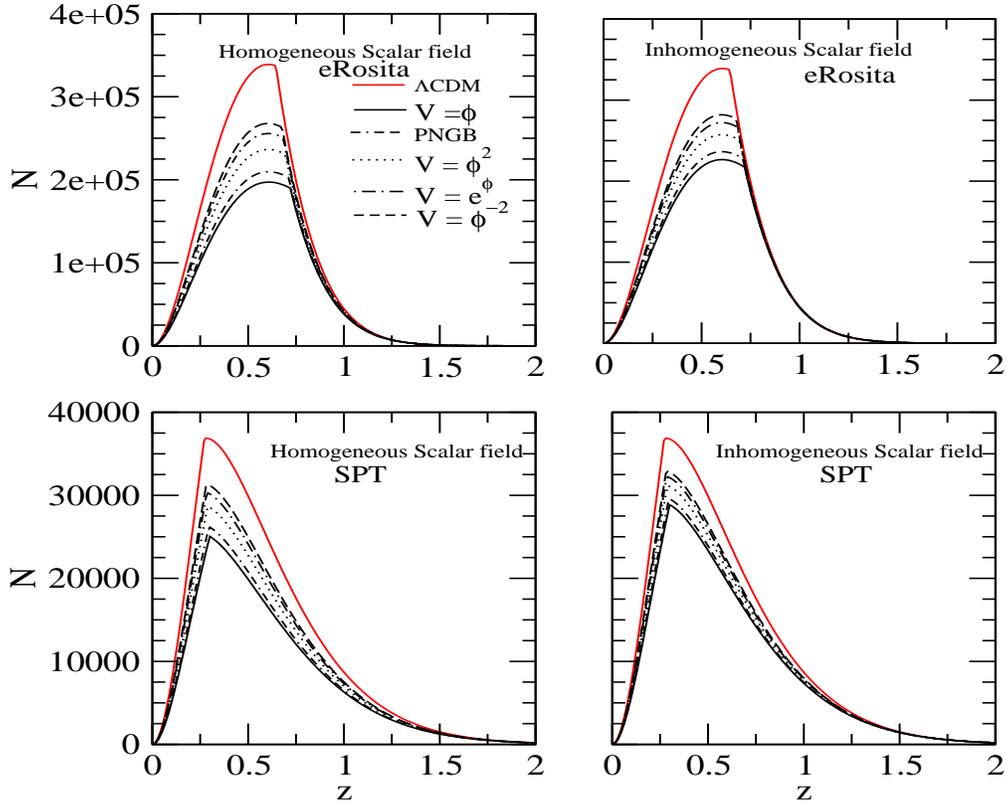}
\caption{The total number counts $N$ as a function of redshift of $z$ for canonical scalar field models as would be observed in surveys like eROSITA (top panels) and SPT (bottom panels). The left panels correspond to canonical scalar field models with homogeneous dark energy while the right panels are for inhomogeneous dark energy models. The different potentials correspond to the different line types. The concordance $\Lambda$CDM model (Bottom Solid line) is also plotted for comparison. The cosmological 
parameters are taken to be
$\Omega_{m0}=0.25,~
\Omega_{\phi0}=0.75,~
h = 0.702,~
\Omega_{B0} = 0.0456~{\rm and}~
n_s= 0.968$. The normalization of the power spectrum is mentioned in the text.}
\label{fig5:Nz}
\end{figure*}

\begin{figure*}
\vspace{1.5cm}
\includegraphics[width=14cm,height=12cm]{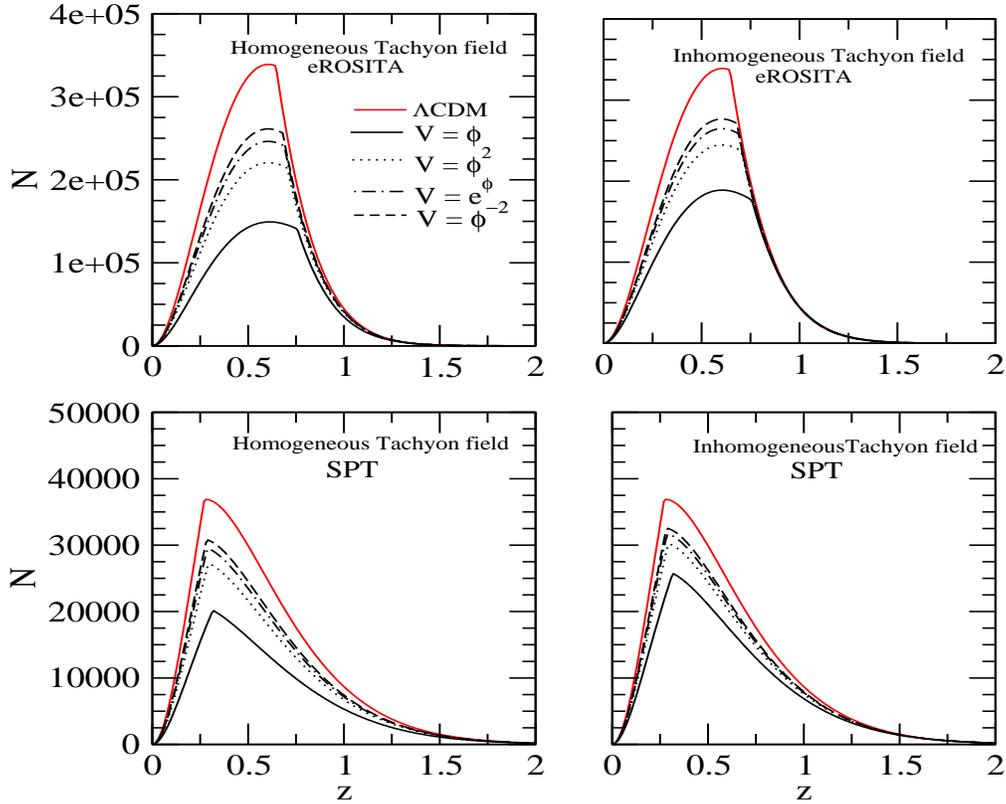}
\caption{Same as Figure \ref{fig5:Nz} but for tachyon scalar fields.}
\label{fig5:Nztach}
\end{figure*}

In passing, we would like to mention that one can find a fitting formula of the linear density contrast for the different models like \citet{Abramo2007} as a function of $\Omega_{m0}$ and redshift. Following \citet{Abramo2007, Weinberg2003, Shaw2008}, the fitting formula for the scaler field with canonical kinetic term has the form of
\beq
\delta_{c}(z)= \frac{3}{20}\times (12\pi)^{\frac{2}{3}}\big(a_{1} + b w(z) +c {\rm log}_{10}[ \Omega_{m}(z)]\big)
\label{fitdeltac}
\eeq
The values of parameters $a_{1}, b$ and $ c$ are different according to the model. Considering the canonical scalar field model, we have listed out the corresponding values in the Table \ref{tab:fitting}. One can also find the similar kind of the fitting formula for other models. We compare the accuracy of this fitting formula plotting with the actual data in Fig \ref{fig:fitting} and one can easily see that the match is excellent.

Next, we investigate a quantity closely related to observations, the clusters number density, $dn/d{\rm log}M$. Since it is only depends on $\delta_c$ and on the growth factor, no appreciable differences are expected between the models studied. In order to see the significant deviation in the cluster number density for the thawing models from that of the standard $\Lambda$CDM, we define a new parameter, $\alpha$ such that 
\begin{equation}
10^{\alpha}=\frac{dn/d{\rm log}M}{\left(dn/d{\rm log}M\right)_{\Lambda CDM}}.
\label{eq:alpha}
\end{equation}

 This parameter measures the deviation in the clusters number density of any thawing models from that of $\Lambda$CDM. Larger the value of $\alpha$ more it deviates from $\Lambda$CDM. The behavior of $\alpha$ for the ordinary scalar field  as well as for the tachyon type field have been shown in Figs.(\ref{fig3:diffscalar}) and (\ref{fig4:difftach}) respectively.
It can be seen from Fig.(\ref{fig3:diffscalar}) that $\alpha$ value is lower for higher $z$, i.e., the clusters number density for the object collapsing earlier have significant deviation from that of  $\Lambda$CDM. This is true for both ordinary scalar field and tachyon field in homogeneous as well as inhomogeneous DE cases. The difference between the homogeneous and inhomogeneous is prominent for the object collapsing at the later time (i.e. at redshift $z = 0.5$).
 Again if we look further, the tachyon field models show larger deviations from the $\Lambda$CDM at all redshifts than the scalar field with canonical kinetic term. Among the potentials we considered, the linear potential again shows the maximum deviation for the tachyon dark energy models for homogeneous as well as inhomogeneous cases.

An  another important quantity that can be derived from observations is the total number 
counts of CDM halos above a given mass in a complete survey volume.
There are a number of cluster surveys ongoing or 
being planned in near future, e.g., 
PLANCK, eROSITA, WFXT
which would detect a large number of clusters \citep{Vikhlinin2009b}. 
To obtain the 
mass of the clusters detected, one usually utilizes 
a proxy variable which can be
the X-ray flux, SZE flux or richness of the cluster. The actual quantity
chosen would, of course, depend on the nature of the survey. The relation
between these quantities and the mass depends on the detailed
cluster physics. However, for our purpose, it suffices to state
that the limiting mass $M_{\rm min}(z)$ of the survey will be essentially
determined by the limits on the proxy variable.

For predicting the number of clusters that would be detected in future
surveys, we choose two specific surveys, namely, 
(i) the eROSITA
satellite X-ray survey, which is expected to have a flux limit 
$f_{\rm lim} = 3.3 \times 10^{-14}$ ergs s$^{-1}$ cm$^{-2}$  at the energy band 0.5-5 keV and would
cover $\sim 20000$ deg$^2$ of the sky (corresponding to a fraction
$f_{\rm sky} \approx 0.485$), and (ii) the South
Pole Telescope SZ survey (SPT), which has a limiting flux density 
$f_{\nu_0, {\rm lim}} = 5$ mJy at the frequency $\nu_0 = 150$ GHz 
and a sky coverage of
$\sim 4000$ deg$^2$ (corresponding to a fraction
$f_{\rm sky} \approx 0.097$). The method of converting these limiting
fluxes to limiting halo masses for the two surveys 
is outlined in \cite{Fedeli2009} and \cite{basik2010}.
We simply follow their prescription and refer the reader to the above papers for
details.

We show the calculated value of the total number counts of haloes 
as the function of redshift in Figures \ref{fig5:Nz} and \ref{fig5:Nztach}
 for all the models 
we considered as indicated in the figures. The top panels correspond to
our predictions for eROSITA while the bottom panels are for SPT. 
We discuss the results below.

It is clear that there exists significant difference in the
number counts between different DE models
considered. The difference is most significant around
$z \sim 0.5$. To take a specific example, 
we see from the top-left panel of Figure \ref{fig5:Nz} that the difference 
in number counts (as would be observed in eROSITA)
between $\Lambda$CDM and $V=\phi$ homogeneous scalar field model is
$\sim 2 \times 10^5$ at $z \sim 0.5$. This difference is significantly
larger than the statistical uncertainties (which would be
$\sim 500$ for the survey we are considering), and hence
can be used for discriminating between different models. 
The corresponding difference in the two models for SPT is 
$\sim 10000$ at $z \sim 0.5$ (see the bottom-left panel), while the
corresponding statistical uncertainties would be $\sim 150$. Hence, the 
two surveys could, in principle, be able to distinguish between $\Lambda$CDM
and the other DE models.
The same conclusions can be drawn from other panels as well, and also
for tachyonic fields shown in Figure \ref{fig5:Nztach}.
We should mention that discriminating between different DE models
would be limited by our ignorance of other cosmological parameters
like $\Omega_m, \sigma_8, n_s$. A proper analysis of 
how to constrain the models would involve error estimates of the
parameters. In the next section, we make a brief attempt in this regard.
 
\section{\normalsize{Observational Constraints}}

\subsection{Cluster Data}

\begin{table*}
\vspace*{-0.3cm}
\caption{Redshift intervals of the massive X-ray cluster data.
 $f_{sky}(i)$ represents the
effective fraction of the observed comoving volume of the
$i^{\rm th}$ bin, \citet{Campanelli2011}.}

\vspace{0.5cm}

\begin{tabular}{ccccccc}

\hline \hline

&bin $i$  &$z_1^{(i)}$  &$z_2^{(i)}$  &$z_c^{(i)}$  &Ref.  &$f_{sky}(i)$ \\

\hline

&1  &0.00  &0.10 &0.050 &\cite{Ikebe2002}    &0.309 \\
&2  &0.30  &0.50 &0.375 &\cite{Henry2000}    &0.012 \\
&3  &0.50  &0.65 &0.550 &\cite{Bahcall1998}, &0.006 \\
&    &      &     &      &\cite{Bahcall2003b} &     \\
&4  &0.65  &0.90 &0.825 &\cite{Donahue1998} &0.001 \\

\hline \hline

\end{tabular}
\label{tab:rebshift}
\end{table*}

\begin{table*}
\caption{ Observational data for Cluster Number Counts for different redshifts bins obtained from the Massive X-ray clusters, \citet{Campanelli2011}.}

\vspace{0.5cm}

\begin{tabular}{llcccc}

\hline \hline

&   &bin 1  &bin 2  &bin 3  &bin 4 \\

&   &$[T_{X,0}(\keV) \, , \mathcal{N}_{{\rm obs},1}]$  &$[T_{X,0}(\keV) \, , \mathcal{N}_{{\rm obs},2}]$
    &$[T_{X,0}(\keV) \, , \mathcal{N}_{{\rm obs},3}]$  &$[T_{X,0}(\keV) \, , \mathcal{N}_{{\rm obs},4}]$ \\

\hline

&$\Delta_v' \in    [25,175]$     &$[7.37\, , 5^{+1}_{-0}]$  &$[9.6\, ,0^{+0}_{-0}]$  &$[10.9\, ,0^{+1}_{-0}]$  &[$12.8\, ,0^{+1}_{-0}]$  \\
&$\Delta_v' \in \; ]175,375]$    &$[6.15\, ,15^{+2}_{-4}]$  &$[8.1\, ,1^{+0}_{-1}]$  &$[9.1 \, ,1^{+1}_{-0}]$  &[$10.7\, ,1^{+0}_{-0}]$  \\
&$\Delta_v' \in \; ]375,750]$    &$[5.54\, ,21^{+2}_{-5}]$  &$[7.3\, ,1^{+4}_{-1}]$  &$[8.2 \, ,2^{+0}_{-1}]$  &[$9.6 \, ,1^{+0}_{-0}]$  \\
&$\Delta_v' \in \; ]750,1750]$   &$[5.14\, ,24^{+1}_{-1}]$  &$[6.7\, ,2^{+4}_{-1}]$  &$[7.6 \, ,2^{+0}_{-1}]$  &[$8.9 \, ,1^{+0}_{-0}]$  \\
&$\Delta_v' \in \; ]1750,3250]$  &$[4.91\, ,24^{+2}_{-0}]$  &$[6.4\, ,5^{+1}_{-3}]$  &$[7.3 \, ,2^{+0}_{-0}]$  &[$8.5 \, ,1^{+0}_{-0}]$  \\

\hline \hline

\end{tabular}
\label{tab:data}
\end{table*}


In this section, we use observational data for cluster abundances to constrain the thawing class of models that behave significantly different from $\Lambda$CDM model.

In this regard we follow the recent treatment by  \cite{Campanelli2011} to calculate the observed galaxy number counts from the X-Ray temperature measurements for massive clusters and the subsequent statistical analysis.

The comoving number for the clusters within the redshift range between $z_{1} $ and $z_{2}$,  with mass $M$ greater than $M_{1}$ is given by \cite{Campanelli2011}:

\begin{equation}
\label{NumberP}
\mathcal{N} = \int_{z_1}^{z_2} \! dz \, \frac{d [f_{sky}(z) V(z)]}{dz} \, N(M>M_{1},z) \ ,
\end{equation}
where $N(M>M_{1},z)$  is the comoving cluster number density
at redshift $z$ for those clusters which have  masses $M$ greater than $M_{1}$ and is given by equation (\ref{eq:dNdz}).  Following \cite{Bahcall1998} and \cite{Bahcall2003b}, we assume $M_{1} = 8 \times 10^{14} h^{-1} M_{\odot}$ with corresponding comoving radius $R_{1} = 1.5 h^{-1}Mpc$. We assume $M_{\odot} \sim 1.989 \times 10^{33} gms $ for the solar mass. As mention earlier
$f_{sky}(z)$ represents the effective fraction of the total comoving volume that is observed at redshift $z$.

Subsequently, we show in Table~\ref{tab:rebshift}, four redshift bins centered at some redshift $z_{c}$. This has been taken from the analysis by  \cite{Campanelli2011} (See also the corresponding references). In the table, the value of the effective fraction $f_{sky}$ of the total comoving volume  are also listed which have been computed using the results of \citep{Bahcall1998,Bahcall2003b}. Usually the $f_{sky}$ parameter depends on the underlying cosmology but this dependence is weak as observed in \cite{Campanelli2011}. Subsequently for i-th bin, we write 

\begin{equation}
\label{NExpected}
\mathcal{N}_i = f_{sky}(i) \int_{z_1^{(i)}}^{z_2^{(i)}} \ dz \, \frac{dV(z)}{dz} \, N(M>M_{1},z) \ ,
\end{equation}
where  $f_{sky}(i)$ is the effective fraction in the $i^{\rm th}$ bin listed in Table~\ref{tab:rebshift}.

Next one uses the X-ray temperature measurements for massive clusters and then convert the temperature measurements to the mass measurements. The detail of this procedure is given in the recent analysis by \cite{Campanelli2011}. We skip this detailing 
and quote the corresponding data for observed number of clusters in Table~\ref{tab:data} as obtained by \cite{Campanelli2011}. 

The quantity $\Delta'_v$ in Table~\ref{tab:data} is defined as:

\begin{equation}
\label{DeltaDelta}
\Delta'_v = \frac{\Omega_m (1+z)^3}{E^2(z)} \, \Delta_v \ .
\end{equation}
The quantity $\Delta_v$ is the virial over-density which depends on redshift and cosmology. For our case it has been thoroughly
discussed in \citet{Devi2011} and in Appendix C of \cite{Campanelli2011}.

\subsection{Data Analysis and Results}

Due to a small data set, one can do the data analysis assuming the error distribution to follow a Poisson statistics. Following \cite{Campanelli2011}, we write the $\chi^{2}$ in our case as:

\begin{eqnarray}
\label{chi2N-nosys}
&& \!\!\!\!\!
\chi^2(n_{s},\Omega_{m0},\sigma_8, \lambda_{i}, \Gamma) =  -2\ln \mathcal{L} \\
&& \simeq 2\sum_{i=1}^4 \left[ \mathcal{N}_i - {\mathcal{N}_{{\rm obs},i}}
\left(1 + \ln \mathcal{N}_i - \ln {\mathcal{N}_{{\rm obs},i}} \right) \right] \ . \nonumber
\end{eqnarray}
Following \cite{Campanelli2011} we also take into account the uncertainty in the comoving numbers of clusters,
$\Delta {\mathcal{N}_{{\rm obs},i}}$, by introducing another parameter $\xi$ to modify the $\chi^2$ as
\begin{eqnarray}
\label{chi2N}
&& \!\!\!\!\!
\chi^2(n_{s},\Omega_{m0},\sigma_8,\lambda_{i},\Gamma,\xi) = \\
&& = 2\sum_{i=1}^4 \left[ \mathcal{N}_i - {\mathcal{N}'_{{\rm obs},i}}
\left(1 + \ln \mathcal{N}_i - \ln {\mathcal{N}'_{{\rm obs},i}} \right) \right] + \xi^2 \ . \nonumber
\end{eqnarray}
where $\mathcal{N}'_{obs,i}=\mathcal{N}_{obs,i}+\xi \Delta\mathcal{N}_{obs,i}$.

As we are interested in thawing models which differ substantially from $\Lambda$CDM behaviour, we fix $\lambda_{i}=1$. We also fix $h=0.72$ for our subsequent analysis. In this present analysis, we consider the thawing models with canonical scalar fields and consider only the inhomogeneous scalar field case.
 \begin{figure}
\includegraphics[width=80mm]{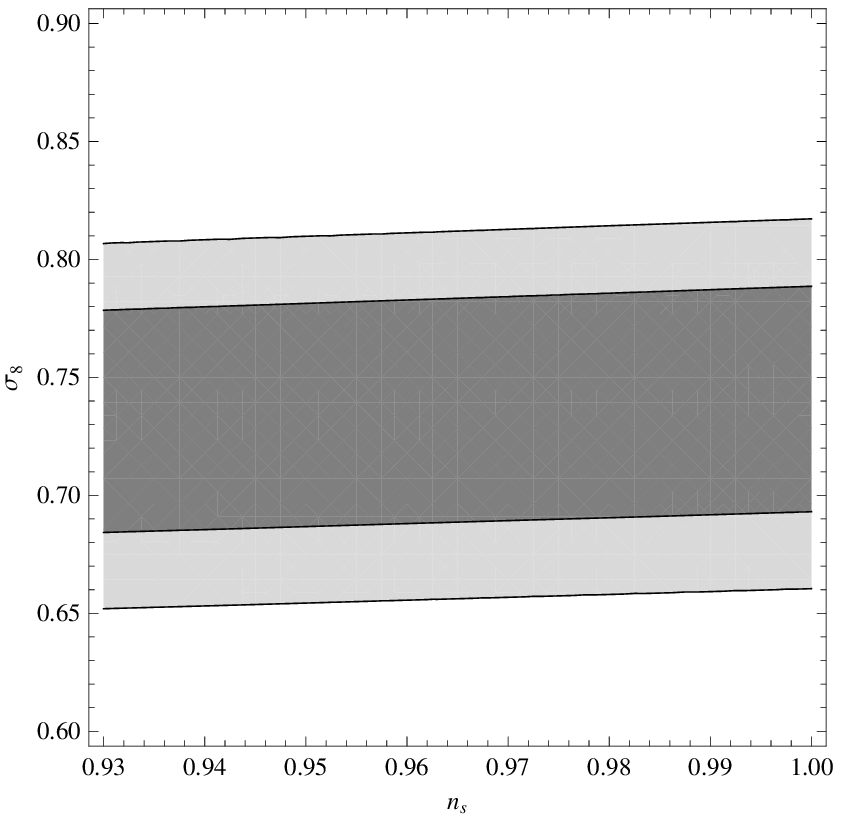}
\caption{$1\sigma$ and $2\sigma$ confidence level contours in the $n_{s}-\sigma_{8}$ plane for the linear potential.}
\label{fig6:chins}
\end{figure}

\begin{figure}
\includegraphics[width=80mm]{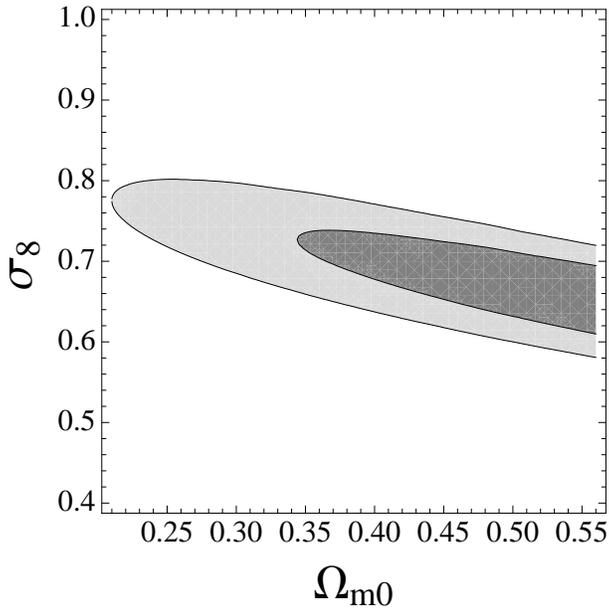}
\caption{$1\sigma$ and $2\sigma$ confidence level contours in the $\sigma_{8}-\Omega_{m0}$ plane.}
\label{fig7:chis8om}
\end{figure}

\begin{figure}
\includegraphics[width=80mm]{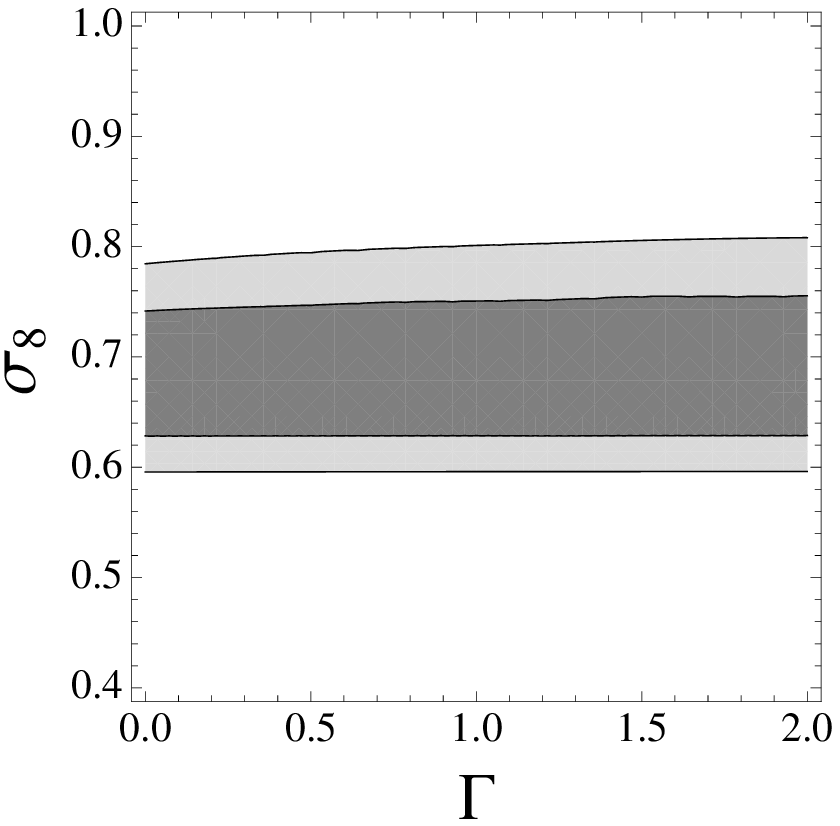}
\caption{$1\sigma$ and $2\sigma$ confidence level contours in the $\sigma_{8}-\Gamma$ plane. after marginalizing over the other parameters.}
\label{fig8:chis8}
\end{figure}  

We marginalise over the parameter $\xi$ with uniform prior. With this, the $\chi^2$ is a function of four parameters, e.g $n_{s}, \Omega_{m0}, \sigma_{8}$ and $\Gamma$. The last one is a parameter related to the scalar field thawing model. It determines the form of the potential. For $\Gamma=0$ i.e for the linear potential,  we show the confidence contours in the $n_{s}-\sigma_{8}$ plane in figure \ref{fig6:chins}. This shows that there is no bound on $n_{s}$. In other words, the cluster data that we have used, can not constrain $n_{s}$. This is also true for other potentials. Hence in our subsequent analysis we fix $n_{s}=0.968$ which is the best fit value obtained by WMAP-7, \cite{Komatsu2011}.

We are now left with three parameters: two cosmological parameters, $\sigma_{8}$ and $\Omega_{m0}$ and one model parameter $\Gamma$.  The results are shown in figures \ref{fig7:chis8om} and \ref{fig8:chis8}. From figure\ref{fig7:chis8om}, it is clear that we get a bound in the $\sigma_{8}-\Omega_{m0}$ plane which is similar to bounds obtained in other works \cite{Campanelli2011}. On the other hand, figure \ref{fig8:chis8} implies that although one can get a strong constraint on the parameter $\sigma_{8}$, it is not possible to constrain the form of the potential for the thawing models as the parameter $\Gamma$ is unconstrained. This is hardly surprising as we have already seen that the data on cluster counts can distinguish between thawing models mainly in the redshift range $0.5-1.0$. However, the data set used in this analysis contains only three clusters in this redshift range. Using the parameters appropriate for these observations, we find that the difference in the cluster counts for various models is of the order of a few, which is similar to the statistical uncertainties. In other words, we can not constrain the thawing dark energy potentials with such a small number of clusters at $z \sim 0.5-1.0$. As we have mentioned earlier, large cluster surveys with substantially more number counts ($ \sim 10^4 $) will be useful in constraining the models.

\section{Conclusion}
To summarize, we  investigate the cluster abundance in cosmological scenario where the universe is dominated by  the thawing class of scalar dark energy. We consider both the ordinary
 scalar field with canonical kinetic term as well as  tachyon field with DBI form of kinetic energy. Moreover, we consider a variety of potentials that can give rise to suitable cosmological scenario different from concordance $\Lambda$CDM model. To study the formation of collapsed structures, we consider homogeneous dark energy as well as dark energy scenario where dark energy takes part in the virialisation process. We consider the Press-Schetcher formalism modified by \citet{Sheth1999} to calculate the mass function. Subsequently, we show that there exists significant difference in the number counts between different dark energy models and the concordance $\Lambda$CDM model. We also constrain our model using presently available observational data for cluster number counts. We show that although the present data is not suitable enough to distinguish thawing class of models, one can still get a reasonably strong constraint on the cosmological parameter $\sigma_{8}$ as well as  a lower bound on the parameter $\Omega_{m0}$.
 
 Given the fact that a large number of cluster surveys are currently ongoing as well as a number of future surveys are being planned, cluster number counts can be a smoking gun to distinguish different dark energy models from the $\Lambda$CDM. Our present work is one preliminary step towards that direction. This work can be extended to freezing class of scalar field models as well as for models with non minimally coupled scalar fields. This will be our future aim.

 \section{Acknowledgement}
AAS acknowledges the financial support provided by the SERC, DST,
Govt. of India through the research grant (DST-SR/S2/HEP-043/2009. NCD
acknowledges the financial support provided by CSIR, Govt. of India.
NCD also acknowledges the Harish-Chandra Research Institute, Allahabad,
India for  hospitality provided  during her visit  where part of the work has been done. We thank the anonymous referee for his/her comments which helped
us in improving the paper.


\begin{thebibliography}{99}

\bibitem[\protect\citeauthoryear{Abramo et al.}{2007}]{Abramo2007}
Abramo L. R., Batista  R. C., Liberato L. \&  Rosenfeld R., 2007,  JCAP, 0711, 012

\bibitem[\protect\citeauthoryear{Ali et al.}{2009}]{Amna2009}
Ali~A.,~Sami~M.,~\& Sen A.~A., 2009, Phys.\ Rev.\ D, 79, 123501



\bibitem[\protect\citeauthoryear{Bahcall et al.}{2003}]{Bahcall2003a}
Bahcall N. A., et al., 2003, ApJ., 585, 182 

\bibitem[\protect\citeauthoryear{Bahcall \& Fan}{1998}]{Bahcall1998}
Bahcall N. A. \&  Fan X.\, 1998, ApJ., 504, 1

\bibitem[\protect\citeauthoryear{Bahcall \& Bode}{2003}]{Bahcall2003b}
Bahcall N. A. \& Bode P.\, 2003, ApJ, 588, L1
                          
                                           
                         
\bibitem[\protect\citeauthoryear{Baldi et al.}{2010}]{Baldi2010}
Baldi M., Pettorino V., Robbers G. \& Springel V., 2010, MNRAS, 403, 1684



\bibitem[\protect\citeauthoryear{Bartelmann et al.}{1998}]{Bartelmann1998}
Bartelmann M., Huss  A., Colberg  J. M., Jenkins  A.
\&  Pearce F. R., 1998, Astron. Astrophys., 330, 1

\bibitem[\protect\citeauthoryear{Basilakos et al.}{2010}]{basik2010}
 Basilakos S., Plionis M. \& Lima J.~A.~S.,\ 2010, Phys. Rev. D, 82, 083517 

\bibitem[\protect\citeauthoryear{Basilakos et al.}{2009}]{basik2009}
 Basilakos S., Sanchez J.~C.~B. \& Perivolaropoulos L.,\ 2009, Phys. Rev. D, 80, 043530



\bibitem[\protect\citeauthoryear{Basilakos \&  Voglis}{2007}]
{Basilakos2007}Basilakos~S. ~\& Voglis N., 2007, MNRAS, 374, 269, 



\bibitem[\protect\citeauthoryear{Bhattacharya et al.}{2011}]{Bhattacharya2011}
Bhattacharya S., 
Heitmann K., White M., et al.,\ 2011, ApJ., 732, 122



\bibitem[\protect\citeauthoryear{Borgani et al.}{2001}]{Borgani2001}
Borgani S., et al., 2001, ApJ., 561, 13


\bibitem[\protect\citeauthoryear{Caldwell \& Linder}{2005}]{Caldwell2005}
Caldwell~R. R. ~\& Linder~E.~V., 2005, Phys. Rev. Lett., 95, 141301

\bibitem[\protect\citeauthoryear{Campanelli et al.}{2011}]{Campanelli2011}
Campanelli~L., Fogli G. L., Kahniashvili T., Marrone  A.  \& Bharat Ratra, 2011, [arXiv:1110.2310] 

\bibitem[\protect\citeauthoryear{Carroll}{2001}]{Carroll2001}
Carroll S. M., 2001, Living Rev. Rel.,  4, 1 
 
\bibitem[\protect\citeauthoryear{Chevallier \& Polarski}{2001}]{Chevallier2001}
Chevallier~M.~\& Polarski~D., 2011 Int.~J.~Mod.~Phys.~D, 10, 213

\bibitem[\protect\citeauthoryear{Cole et al.}{2005}]{Cole2005}
Cole~S., et al., 2005, MNRAS, 362, 505




\bibitem[\protect\citeauthoryear{Corless \& King}{2009}]{Corless2009}
Corless  V. L. \& King L. J., 2009, MNRAS, 396, 315 
 
\bibitem[\protect\citeauthoryear{Courtin et al.}{2011}]{Courtin2011}
Courtin J. et al., 2011, MNRAS, 410, 1911

\bibitem[\protect\citeauthoryear{Dahle}{2006}]{Dahle2006}
Dahle  H., 2006, ApJ, 653, 954 

\bibitem[\protect\citeauthoryear{Devi \& Sen}{2011}]{Devi2011}
Devi N. C. \& Sen A. A., 2011, MNRAS, 413, 2371

\bibitem[\protect\citeauthoryear{Donahue et al.}{1998}]{Donahue1998}
Donahue M.\,  et al.,\ 1998, ApJ, 502, 550

\bibitem[\protect\citeauthoryear{Eisenstein \& Hu}{1998a}]{Eisenstein1998a}
Eisenstein D. J. \& Hu W., 1998a, ApJ, 496, 605



\bibitem[\protect\citeauthoryear{Evrard et al.}{2002}]{Evrard2002}
Evrard A. E., et al., 2002, ApJ.,  573, 7

\bibitem[\protect\citeauthoryear{Fedeli, Moscardini \& Matarrese}{2009}]{Fedeli2009} 
Fedeli C., Moscardini L. \&  Matarrese S., 2009, MNRAS, 397, 1125 

\bibitem[\protect\citeauthoryear{Francis et al.}{2009}]{Francis2009} 
Francis M. J., Lewis G. F. \&  Linder G. F., 2009, MNRAS, 393, L31 

\bibitem[\protect\citeauthoryear{Frieman et al.}{1995}]{Frieman1995}
Frieman J. A., Hill C. T., Stebbins A. \& Waga I., 1995, Phys. Rev. Lett., 75, 2077

\bibitem[\protect\citeauthoryear{Garousi}{2000}]{Garousi2000}
Garousi M. R., 2000, Nuclear Phys. B, 584, 284


\bibitem[\protect\citeauthoryear{Gunn \& Gott}{1972}]{Gunn1972}
Gunn J. E. \& Gott J. R., 1972, ApJ, 176, 1

\bibitem[\protect\citeauthoryear{Gupta, Majumdar \& Sen}{2012}]{Gupta2012}
Gupta G., Majumdar S. \& Sen, A.~A., 2012, MNRAS, 420, 1309

\bibitem[\protect\citeauthoryear{Henry}{2000}]{Henry2000}
 Henry J.\ P.\, 2000, ApJ, 534, 565


\bibitem[\protect\citeauthoryear{Ikebe et al.}{2002}]{Ikebe2002}
Ikebe  Y.\, Reiprich T.\ H.\, Boehringer H.\, Tanaka Y.\  \& Kitayama T.\, 2002, Astron.\ Astrophys.\, 383, 773
                            

                            
\bibitem[\protect\citeauthoryear{Kluson}{2000}]{Kluson2000}
Kluson J., 2000, Phys. Rev. D, 62, 126003

\bibitem[\protect\citeauthoryear{Khedekar et al.}{2010b}]{Khedekar2010a}
Khedekar S., Das  S. \& Majumdar S., 2010a, Phys. Rev. D, 82, 1301

\bibitem[\protect\citeauthoryear{Khedekar \& Majumdar}{2010b}]{Khedekar2010b}
Khedekar S. \& Majumdar S., 2010b, Phys. Rev. D, 82, 081301 

\bibitem[\protect\citeauthoryear{Komatsu et al.}{2009}]{Komatsu2009}
Komatsu E., et al., 2009, Astrophys. J. Suppl. 180, 330

\bibitem[\protect\citeauthoryear{Komatsu et al.}{2011}]{Komatsu2011}
Komatsu E., et al., 2011, Astrophys. J. Suppl. 192, 18


\bibitem[\protect\citeauthoryear{Kowalski et al.}{2008}]{Kowalski2008}
Kowalski M., et al., 2008, ApJ, 686, 749  

\bibitem[\protect\citeauthoryear{Liberato \& Rosenfeld}{2006}]{Liberto2006}
Liberato~L. \& Rosenfeld~R., 2006, JCAP, 7, 9

\bibitem[\protect\citeauthoryear{Linder}{2003}]{Linder2003}
Linder~E. V., 2003, Phys. Rev. Lett. 90, 091301

\bibitem[\protect\citeauthoryear{Macci et al.}{2004}]{Maccio2004}
Maccio A. V., Quercellini C., Mainini R., Amendola L. \& Bonometto S. A., 2004, Phys. Rev. D, 69, 123516 




\bibitem[\protect\citeauthoryear{ Manera \&  Mota}{2006}]{Manera2006}
Manera~M. \& Mota~ D. F., 2006, MNRAS, 371, 1373

\bibitem[\protect\citeauthoryear{Mota}{2008}]{mota2008}
Mota D.~F.\ 2008, JCAP, 9, 6


\bibitem[\protect\citeauthoryear{Mortonson}{2009}]{Mortonson2009}
Mortonson M. J., 2009, Phys. Rev. D, 80, 123504 

\bibitem[\protect\citeauthoryear{Nunes \& Mota}{2006}]{Nunes2006}
Nunes N. J. \& Mota D. F., 2006, MNRAS, 368, 751

\bibitem[\protect\citeauthoryear{Pace et al.}{2010}]{Pace2010}
Pace  F., Waizmannm  J-C. \& Bartelman  M., 2010, MNRAS,  406, 1865


\bibitem[\protect\citeauthoryear{Percival et al.}{2009}]{Percival2009}
Percival W.~J., et al., 2009, MNRAS, 401, 2331.
 

\bibitem[\protect\citeauthoryear{Press \& Schechter}{1974}]{Press1974}
Press W. H. \& Schechter P., 1974, ApJ, 187, 425

\bibitem[\protect\citeauthoryear{Reed et al.}{2007}]{Reed2007}
Reed D.~S., Bower R., Frenk C.~S., Jenkins A. \& Theuns T.,\ 2007, MNRAS, 374, 2


\bibitem[\protect\citeauthoryear{Reiprich \& Bohringer }{2002}]{Reiprich2002}
Reiprich T. H. \& B¨ohringer H., 2002 ApJ., 567, 716,


\bibitem[\protect\citeauthoryear{Riess et al.}{2009}]{Riess2009}
Riess A. G., et al., 2009, ApJ, 699, 539  

\bibitem[\protect\citeauthoryear{Sahni \&  Starobinsky}{2000}]{Sahni2000}
Sahni V. \&  Starobinsky A., 2000, Int. J. Mod. Phys. D, 9, 373 

\bibitem[\protect\citeauthoryear{Schaefer \&  Koyama}{2004}]
{Schaefer2008}Schaefer~B. M. \&  Koyama~K., 2008, MNRAS, 385, 411 

\bibitem[\protect\citeauthoryear{Scherrer \& Sen}{2008a}]
{Scherrer2008a}~Scherrer~R. J. \& Sen A. A.,~2008a Phys. Rev. D, 77, 083515

\bibitem[\protect\citeauthoryear{Scherrer \& Sen}{2008b}]
{Scherrer2008b}~Scherrer~R. J. \& Sen~A. A.,~2008b Phys. Rev. D, 78, 067303

 
\bibitem[\protect\citeauthoryear{Sen}{2002a}]{Sen2002a}
Sen~A., 2002a, J. High Energy Phys., 0204, 048

\bibitem[\protect\citeauthoryear{Sen}{2002b}]{Sen2002b}
Sen~A., 2002b, Mod. Phys. Lett. A, 17, 1797

\bibitem[\protect\citeauthoryear{Shaw \& Mota}{2008}]{Shaw2008}
Shaw D. J. \& Mota D. F., 2008, ApJS, 174, 277

\bibitem[\protect\citeauthoryear{Sheth \& Tormen}{1999}]{Sheth1999}
Sheth~R. K. \& Tormen~ G., 1999, MNRAS, 308, 119

\bibitem[\protect\citeauthoryear{Tauber}{2005}]{Tauber2005}
Tauber J. A., 2005, New Cosmological Data and the values of
the Fundamental Parameters,  201, 86

\bibitem[\protect\citeauthoryear{Vikhlinin et al.}{2009a}]{Vikhlinin2009a}
Vikhlinin  A., et al., 2009, ApJ., 692, 1060

\bibitem[\protect\citeauthoryear{Vikhlinin et al.}{2009b}]{Vikhlinin2009b}
Vikhlinin  A., et al., 2009b, [arXiv:0903.5320]

\bibitem[\protect\citeauthoryear{Wang \& Steinhardt}{1998}]{Wang1998}
Wang L. \& Steinhardt P. J., 1998, ApJ, 508, 483

\bibitem[\protect\citeauthoryear{Weinberg}{1986}]{Weinberg1986}
Weinberg S., 1986, Rev. Mod. Phys., 61, 1 

\bibitem[\protect\citeauthoryear{Weinberg \& Kamionkowski}{2003}]{Weinberg2003}
Weinberg N. N. \& Kamionkowski M., 2003, ApJ, 341, 251

\bibitem[\protect\citeauthoryear{Wen}{2010}]{Wen2010}
Wen Z. L., Han J.~L. \& Liu  F.~S., 2010, 2010, MNRAS, 407, 533  


\end{thebibliography}
\end{document}